\documentclass[a4paper,11pt]{article}
\usepackage{jinstpub} 
\usepackage{lineno}
\usepackage{hyperref}

\proceeding{7$^{\text{th}}$ International Conference on Micro-Pattern Gaseous Detectors\\
11–16 December 2022\\
Rehovot, Israel}

\title{\boldmath Studying signals in particle detectors with resistive elements such as the 2D resistive strip bulk MicroMegas}

\author[a,b,c,1]{D. Janssens \note{Corresponding author.}}
\author[a]{, F. Brunbauer}
\author[a,d]{, K.J. Flöthner}
\author[a,e]{, M. Lisowska}
\author[a,f]{, H. Muller}
\author[a]{, E. Oliveri}
\author[a,g]{, G. Orlandini}
\author[a]{, W. Riegler}
\author[a]{, L. Ropelewski}
\author[a]{, H. Schindler}
\author[a,f]{, L. Scharenberg}
\author[a,h]{, A. Utrobicic}
\author[a]{and R. Veenhof}

\affiliation[a]{European Organization for Nuclear Research (CERN), 1211 Geneva 23, Switzerland}
\affiliation[d]{Helmholtz-Institut für Strahlen- und Kernphysik, University of Bonn, Nußallee 14–16, 53115 Bonn, Germany}
\affiliation[b]{Inter-University Institute for High Energies (IIHE), Belgium}
\affiliation[c]{Vrije Universiteit Brussel, 1050 Brussels, Belgium}
\affiliation[e]{Université Paris-Saclay, F-91191 Gif-sur-Yvette, France}
 \affiliation[f]{Physikalisches Institut, University of Bonn, Nußallee 12, 53115 Bonn, Germany}
\affiliation[g]{Friedrich-Alexander-Universität Erlangen-Nürnberg, Schloßplatz 4, 91054 Erlangen, Germany}
\affiliation[h]{European Spallation Source ERIC (ESS), Box 176, SE-221 00 Lund, Sweden}
\affiliation[i]{University of Milano-Bicocca, Department of Physics, Piazza della Scienza 3, 20126 Milan, Italy}
\affiliation[h]{Ruder Bošković Institute Bijenička c. 54, 10000 Zagreb, Croatia}

\emailAdd{djunes.janssens@cern.ch}

\abstract{
As demonstrated by the ATLAS New Small Wheel community with their MicroMegas (MM) design, resistive electrodes are now used in different detector types within the Micro Pattern Gaseous Detector family to improve their robustness or performance. The extended form of the Ramo-Shockley theorem for conductive media has been applied to a 1 M$\Omega$/$\Box$ 2D resistive strip bulk MM to calculate the signal's spreading over neighbouring channels using an 80 GeV/c muon track. For this geometry, the dynamic weighting potential was obtained numerically using a finite element solver by applying a junction condition and coordinate scaling technique to accurately represent the boundary conditions of a $10\times 10$ cm$^2$ active area. Using test beam measurements, the results of this model will be used to benchmark this microscopic modelling methodology for signal induction in resistive particle detectors.
}


\begin{document}
\maketitle
\flushbottom

\section{Introduction}\label{sec: Introduction}
Resistive elements are typically implemented in particle detectors for two distinct reasons: (i) to increase their robustness, as is the case with the small-pad resistive MicroMegas (MM) \cite{Iodice2020}, and (ii) to improve their performance, of which the Resistive Silicon Detector is a prime example \cite{Tornago2021}. In the presence of these materials, the total induced signal in the readout electrodes is the sum of the direct induction from the movement of charged particles in the drift medium, and the time-dependent reaction of the resistive materials. For the characterising their response, a growing interest in this class of detectors rises the need of keeping modelling tools such as Garflield++ \cite{Garfield} linked with this development. As a result, these models can inform the design of the next generation of particle detectors driven by the specific needs of future High Energy Physics (HEP) experiments.

\begin{figure}[t!]
\centering 
\includegraphics[width=.4\textwidth]{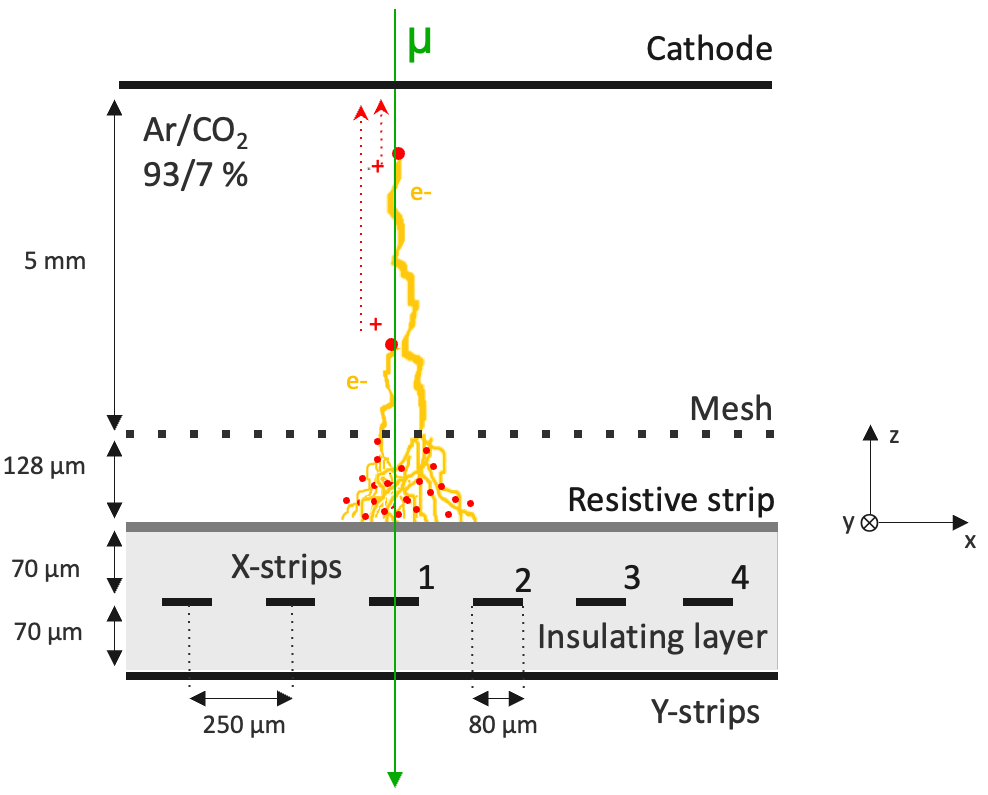}
\qquad
\includegraphics[width=.4\textwidth]{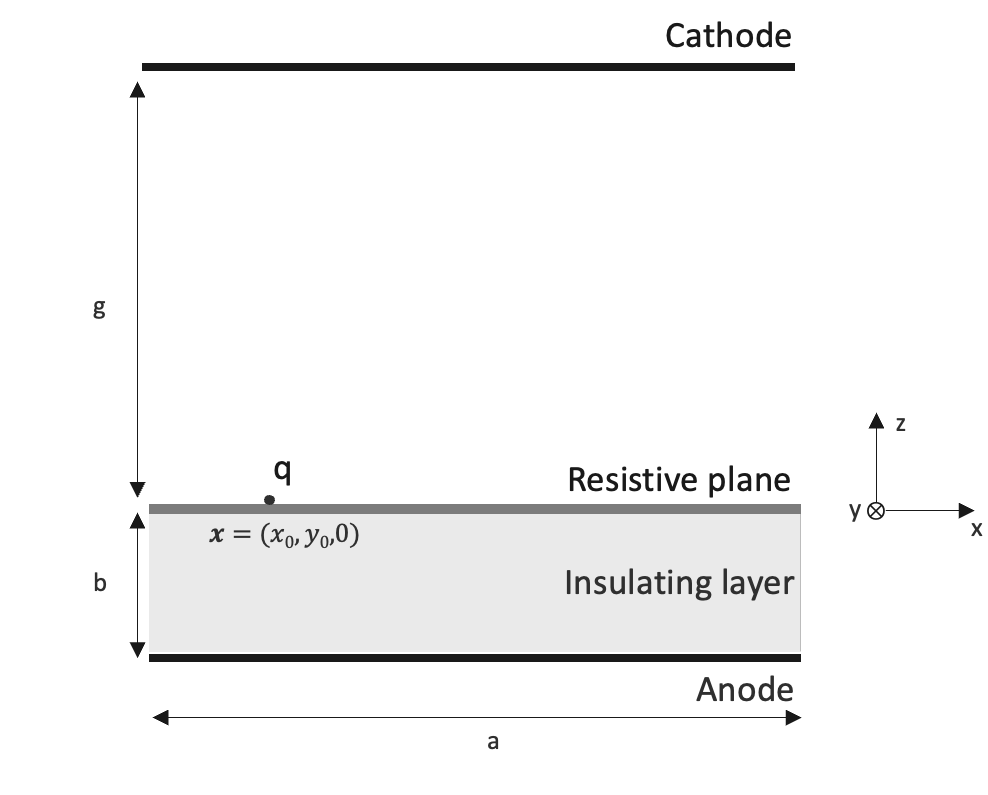}
\caption{\label{fig: Layout of geometry} \textbf{Left:} Schematic overview of the layout of the two-dimensional resistive strip bulk MM. The electron avalanche sourced by a muon going through is sketched in yellow while the positive ion tracks are suppressed to the red arrows. \textbf{Right:} Cross-section view of the toy-model geometry containing a resistive layer of dimensions $a\times a$ that is grounded on all edges with a charge $q$ being deposited on it.}
\end{figure}

The weighting potential becomes time-dependent for geometries containing conductive materials due to the medium’s finite conductivity \cite{Riegler2002,Riegler2004,Riegler2019}. 
Since through analytical methods these potentials can only be obtained for a limited the number of geometries, a numerical method has been suggested to model the time dependence of signals in these readout structures \cite{Janssens2022}. This approach is to be benchmarked for a few key technologies, including the resistive strip bulk MM \cite{Alexopoulos2011}, by comparing with measurements. The resistive strip bulk MM device uses an innovative resistive AC-coupled readout to improve its robustness. For this Micro-Pattern Gaseous Detector (MPGD), it is known that equipped with two-dimensional strip electrodes signal spreading can be observed over neighbouring channels that run perpendicular to the resistive strips \cite{Byszewski2012}. Figure \ref{fig: Layout of geometry} (left) shows a cross-section view of the geometry. This observation spurred the acquisition of two virtually identical prototypes with two different surface resistivities to assemble a benchmarking setup for validating the simulation method using data taken at the H4 extraction beam line of the CERN Super Proton Synchrotron (SPS). Used in the benchmarking effort, this work will describe the ongoing development of the computational model of the response of a 1 M$\Omega/\Box$ 2D resistive strip bulk MM, with a focus on representing the boundary conditions in large-area detectors accurately.

\section{Going to large area detectors}
Since the resistive strips are grounded on one edge of the geometry, the active area's size does affect the evolution of the delayed component of the signal. This impact has been studied analytically by going to a resistive layer MM toy-model before proceeding to the full readout structure.
\subsection{Charge diffusion on a resistive layer}\label{sec: Large area detectors}
\begin{figure}[t!]
\centering 
\includegraphics[width=.3\textwidth]{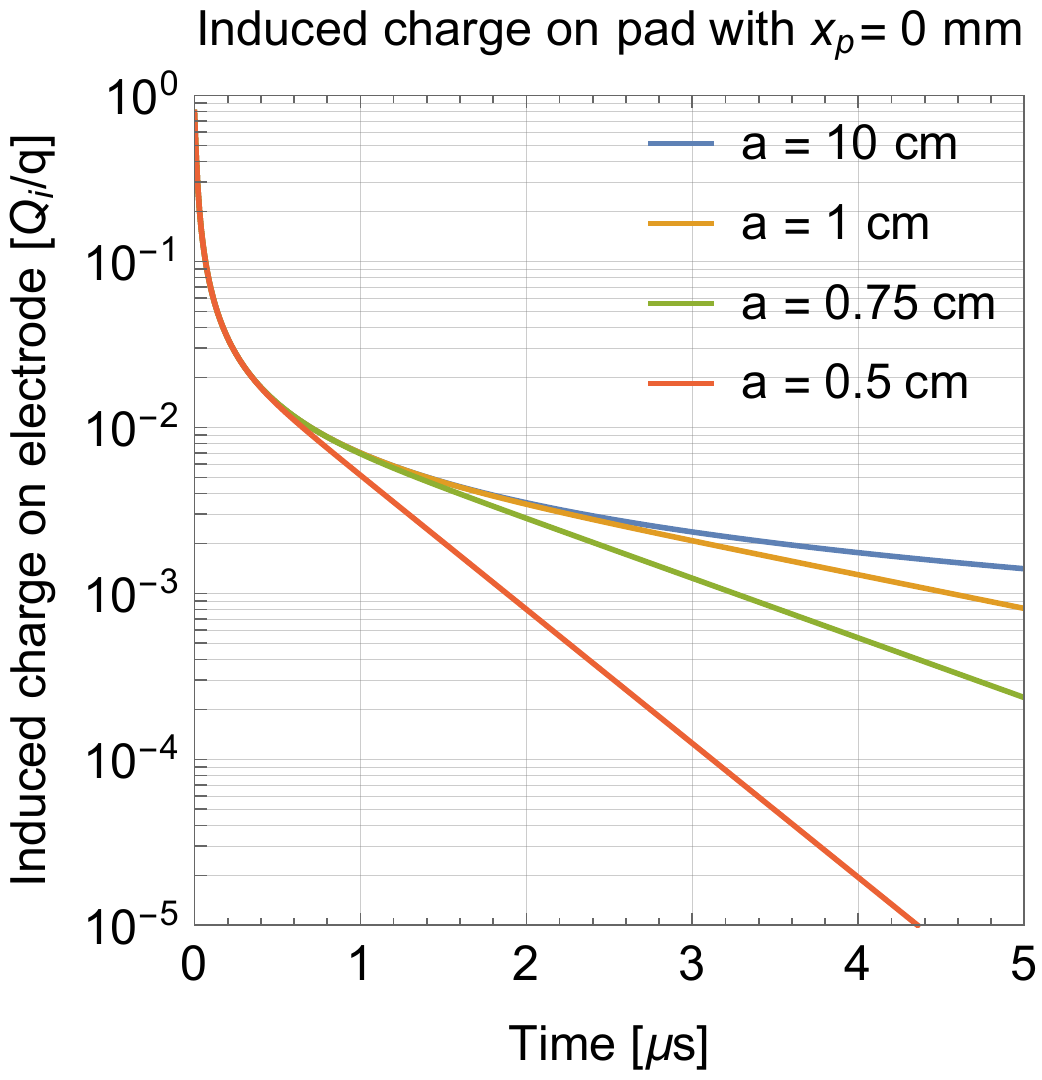}\hfill
\includegraphics[width=.3\textwidth]{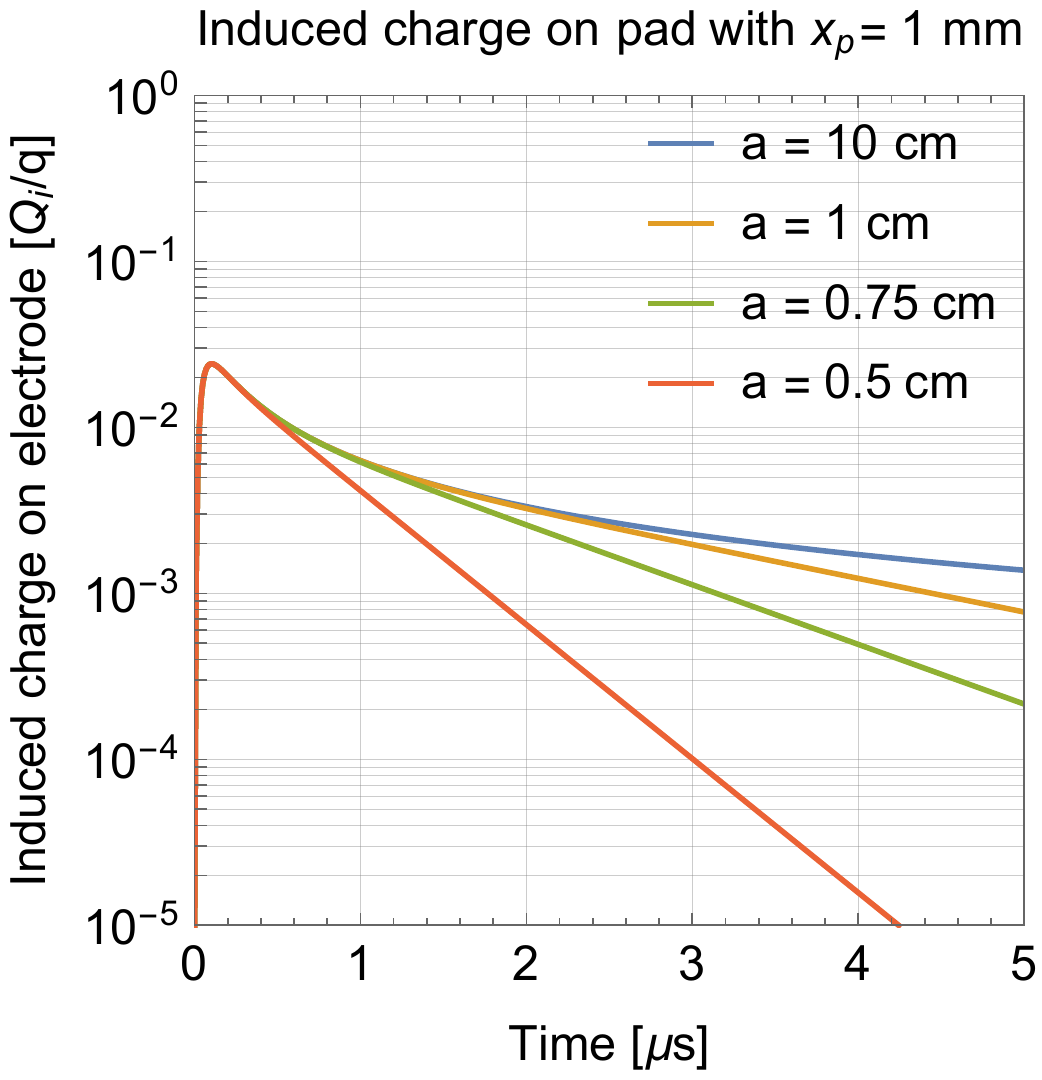}\hfill
\includegraphics[width=.3\textwidth]{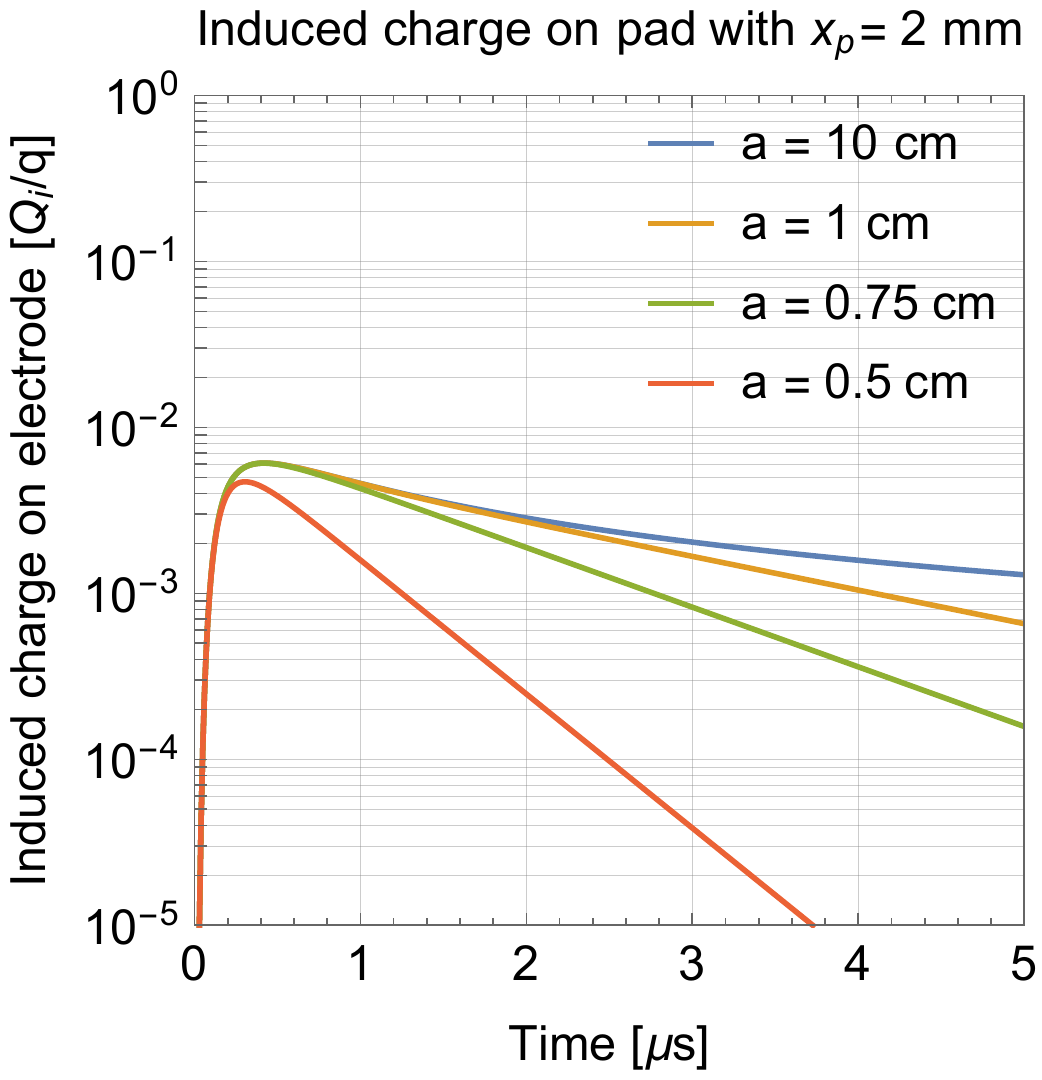}
\caption{\label{fig: Solution electrodes} Induced charge on three $1\times 1$ mm$^2$ neighboring pads positioned at $\mathbf{x}=(x_p,0,-b)$ from a charge $q$ deposited at the center ($\mathbf{x}=\mathbf{0}$) of a square resistive layer with various widths $a$. Here, $g=128$ $\mu$m, $b=100$ $\mu$m, and $R = 1$ M$\Omega/\Box$.}
\end{figure}

For a charge $q$ deposited at time $t=0$ at the centre of a resistive layer grounded on the edge of a finite square parallel plate geometry with length $a$ as depicted in Figure \ref{fig: Layout of geometry} (right), the induced charge on the pads of the segmented anode has been calculated following \cite{Riegler2016}. Using a surface resistivity of $R = 1$ M$\Omega/\Box$ these solutions are shown in Figure \ref{fig: Solution electrodes} for three adjacent pads for different values of $a$. They suggest that in the central region of the detector the early-time behaviour is minimally affected by varying $a$. Nonetheless, at later times, the difference in the size of the resistive layer becomes apparent, where the smaller area detectors drain the diffusing charge distribution in the layer more efficiently. This is reflected in their smaller characteristic time constants of the exponential decaying tail dominating the late-time behaviour. The evaluation of this example can be shown to converge more slowly as the area of the detector increases; requiring a total number of terms proportional to $a^2/b^2$ for $t=0$ alone to be summed up, e.g., $\mathcal{O}(10^7)$ for $a = 10 $ cm. This number quickly rises as soon as one steps into the area of applications in HEP experiments. The convergence speed can possibly be improved by using series acceleration methods. However, for geometries lacking an analytical solution, ways to manage larger area structures numerically need to be devised. 

\subsection{Numerical scaling of the model}\label{sec: Large area detectors}
The large ratio between the detector area's size and the induction region's thickness makes both analytical and finite element numerical evaluations slow and cumbersome. The detector that we aim to describe has a $10 \times 10 $ cm$^2$ active area where the Dirichlet boundary condition of the resistive element needs to be taken into account to give an accurate description. 

To this end, we have used two techniques. The first was to avoid the need to mesh the thin resistive layer using the electric shielding surface condition available in the COMSOL Multiphysics toolkit \cite{COMSOL}. This follows the junction condition 
\begin{equation}
\mathbf{n} \cdot\left(\mathbf{J}_1-\mathbf{J}_2\right)=-\nabla_T \cdot d\left(\left(\sigma+\varepsilon_0 \varepsilon_r \frac{\partial}{\partial t}\right) \nabla_T V\right)\, ,
\end{equation}
for a thin resistive layer with thickness $d$, conductivity $\sigma = (R d)^{-1}$, relative permittivity $\varepsilon_r$, $\mathbf{n}$ the normal vector of the resistive surface, $\mathbf{J}_i$ the current densities of the region above and below the layer and the operator  $\nabla_T$ represents the tangential derivative along the layer. The second point was to stretch the model using coordinate mapping rather than directly implementing the total active area. To accurately represent the boundary conditions of this geometry in the finite element model, linear coordinate scaling was employed outside a sub-region of interest, allowing for a more numerically friendly confined geometrical representation that is smaller than the total detector active area yet equivalent.

\begin{figure}[t!]
\centering 
\includegraphics[width=0.45\textwidth]{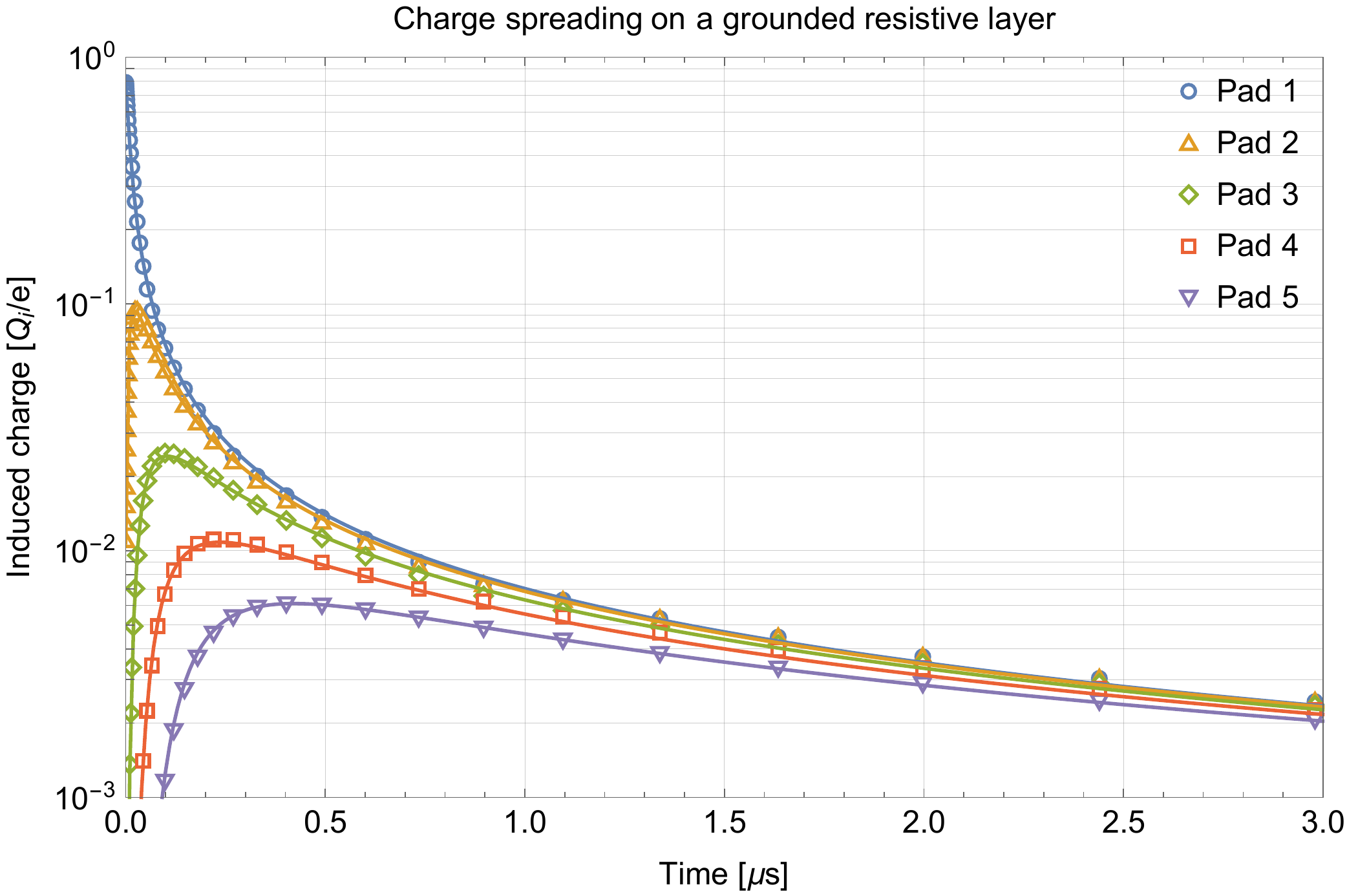}
\caption{\label{fig: comparisonScaling} Comparison between the induced charge on neighbouring pads given analytically and the corresponding COMSOL solution. The former is given by the full lines, while the latter is given by the markers of the same colour for a few time points. The charge is deposited in the centre of a $10 \times 10 $ cm$^2$ resistive layer with R = 1 M$\Omega/\Box$. The five $5 \times 5 $ mm$^2$ pads with index $i$ at $x_p = i-1$ mm.} 
\end{figure}

Going back to the previous example, the time-dependent weighting potential of the pads has been calculated using the finite element method, where these two techniques are applied. The comparison of this solution with the analytical one is shown in Figure \ref{fig: comparisonScaling}, indicating excellent agreement between the two.

\section{Simulations of a resistive strip bulk MM}
To apply the extended form of the Ramo-Shockley (RS) theorem for conductive media numerically to the geometry under investigation, the weighting potential is computed with COMSOL using the techniques described above. Subsequently, we used this solution in Garfield++ to calculate the induced signal on four strips for an event given by a muon track as indicated in Figure \ref{fig: Layout of geometry} (left).
\subsection{Weighting potential solution}\label{sec: Large area detectors}
Three time slices of the weighting potential solution of one of the strips are shown in Figure \ref{fig: results} (left). At $t=0$ the prompt weighting field instantaneously permeates the detector volume, and all elements with finite resistivity appear as insulators. As time increases, the resistive strips diffuse the potential over its surface so that for $t \to \infty$ all conducting materials behave like perfect conductors. 

\subsection{Description of the Monte Carlo model}\label{sec: Large area detectors}
The complete response of the detector is modelled as the result of a series of independent calculations where the MM is approximated as a parallel plate chamber partitioned by an infinitesimally thin metal layer in place of the mesh. The ionisation pattern due to a relativistic muon with a momentum of 80 GeV/c going through the $5$ mm drift gap is given by HEED \cite{Smirnov2005}. For the electrons, the parameters used for their drift, diffusion, amplification and attachment are provided by MAGBOLTZ \cite{MAGBOLTZ} and used to microscopically model their trajectories to the mesh spurred by a 550 V/cm applied electric field. After reaching the mesh, they are transferred to the 128 $\mu$m induction gap where the avalanche dynamics are simulated microscopically given a $510$ V potential difference between the cathode "mesh" substitute and the resistive strips. The resulting trajectories of the positive and negative charge carriers are used in conjunction with the weighting potential solution to calculate the total induced signal on the x-strips. Taken from Garfield++, the ion mobility is scaled up to match an ion tail length measurement performed using a single-channel PICOSEC MM \cite{Bortfeldt2018} non-resistive single-channel prototype using the same gas mixture and equipped with a fast amplifier for the same gas mixture. Finally, the impulse response function of the electronics is convoluted with the total induced signal of the electrodes. We approximate the electronics by an idealized uni-polar shaper:
\begin{equation}\label{Trasfer function}
f(t)=g e^{n}\left(\frac{t}{t_p}\right)^n e^{-\frac{t}{\tau}}\, ,
\end{equation}
where the peaking time is defined as $t_p = n \tau $ \cite{Rolandi2008}. To mimic the response of the APV25 chip we assume $t_p = 50$ ns, and a first order shaping $n = 1$, while keeping the amplification to $g = 1$ \cite{Jones1999,Raymond2000}.

\subsection{Induced signal waveform}\label{sec: Large area detectors}
\begin{figure}[t!]
    \centering
    \vspace*{1in}~\\
        \includegraphics[width=0.51\textwidth]{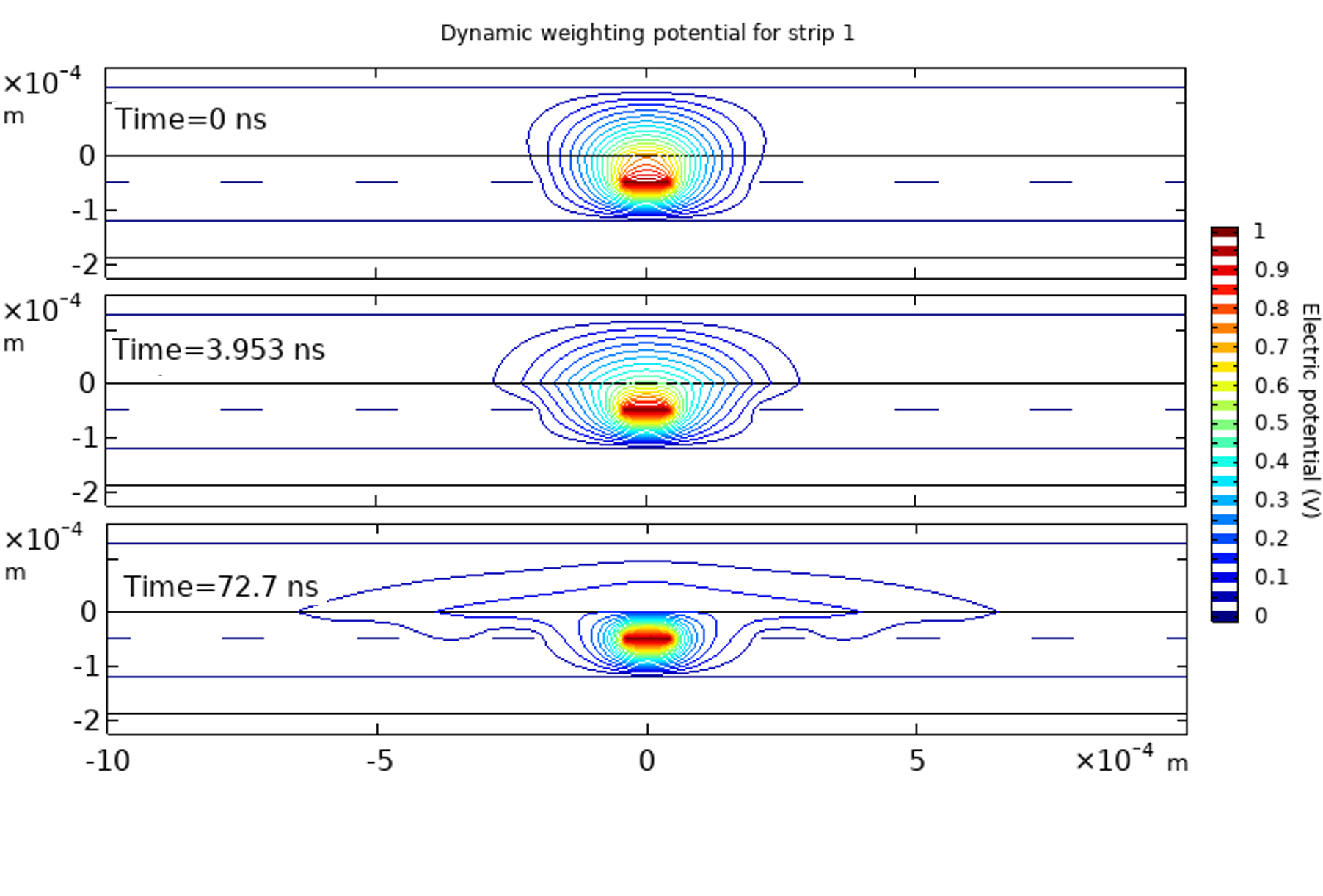}
        \quad
        \includegraphics[width=0.36\textwidth]{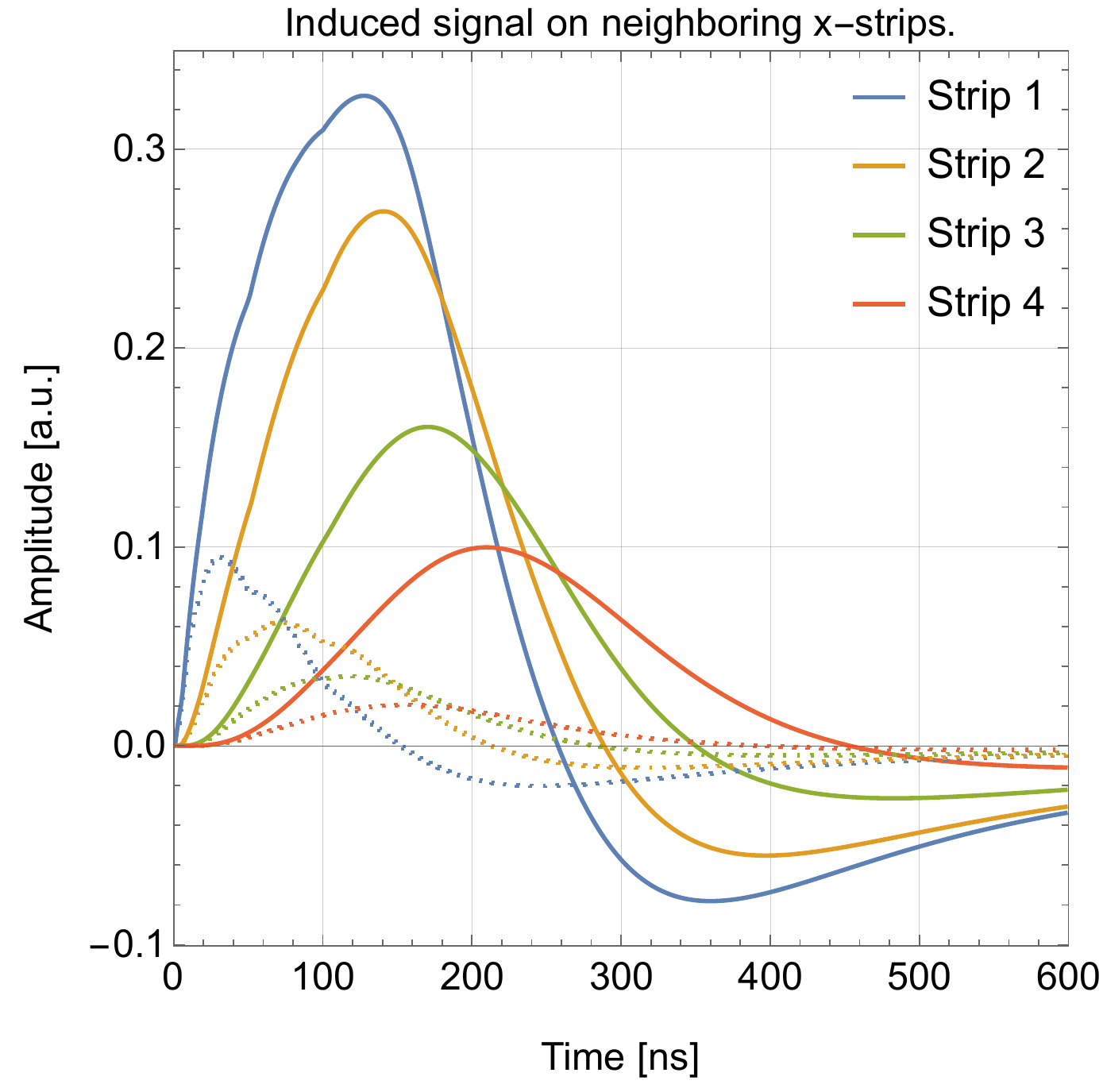}
    \caption[]{\textbf{Left:} Time sliced cross-section weighting potential map for x-strip number 1 with resistive strips of 1 M$\Omega/\Box$. \textbf{Right:} Simulated total induced signals (full lines) and electron contribution (dashed lines) in four neighbouring readout strips normal to the resistive strips with a surface resistivity of $1\,\mathrm{M}\Omega/\Box$ as sourced by a muon perpendicular to the centre of strip 1.}
    \label{fig: results}
\end{figure}

The final bi-polar total induced signal after shaping is shown in Figure \ref{fig: results} (right) alongside its electron component. As expected, the biggest signal amplitude is found in the strip positioned bellow the muon track, whereas the neighbouring channels have progressively smaller peak amplitude values that are shifted in time, indicating the "spreading" of the signal over the resistive strips and resulting in the characteristic v-shape peak position in time curve \cite{Byszewski2012}. The majority of the final waveform is comprised of the ion contribution making the result sensitive to the accuracy of the ion mobility in the gas.

\section{Conclusions and outlook}\label{sec: Conclusions}
In this proceeding, the ongoing development of a computational model for the signal induction in a $1$ M$\Omega/\Box$ 2D resistive strip bulk MM was shown, which will be compared with SPS test beam data. Two techniques were discussed to accurately represent the relatively large areas of the reference devices successfully: (i) the use of a junction condition on the resistive strip boundary, avoiding the need for the meshing of thin layers, and (ii) applying a coordinate mapping to stretch the geometry mathematically. The final induced signals of four readout strips perpendicular to the resistive strips were calculated with Garfield++, which show the expected "spreading" of the signal reported in the literature. Before a detailed comparison with the measured data, refinements of the model in the form of implementation of the mesh structure and adding of capacitive coupling will be explored.

\acknowledgments
We acknowledge the support of the CERN EP R\&D Strategic Programme on Technologies for Future Experiments (\href{https://ep-rnd.web.cern.ch/}{https://ep-rnd.web.cern.ch/}) and the RD51 collaboration (\href{https://rd51-public.web.cern.ch}{https://rd51-public.web.cern.ch}), in the framework of RD51 common projects.


 \bibliographystyle{JHEP}
 \bibliography{biblio.bib}

\providecommand{\href}[2]{#2}\begingroup\raggedright\begin{thebibliography}{10}

\bibitem{Iodice2020}
M.~Iodice et~al., \emph{Small-pad resistive micromegas: Comparison of patterned
  embedded resistors and dlc based spark protection systems.},
  \href{https://doi.org/10.1088/1742-6596/1498/1/012028}{\emph{Journal of
  Physics: Conference Series} {\bfseries 1498} (2020) 012028}.

\bibitem{Tornago2021}
M.~Tornago et~al., \emph{Resistive ac-coupled silicon detectors: Principles of
  operation and first results from a combined analysis of beam test and laser
  data}, \href{https://doi.org/10.1016/J.NIMA.2021.165319}{\emph{Nuclear
  Instruments and Methods in Physics Research Section A: Accelerators,
  Spectrometers, Detectors and Associated Equipment} {\bfseries 1003} (2021)
  165319}.

\bibitem{Garfield}
H.~Schindler and R.~Veenhof, \emph{{Garfield++}},
  {\emph{\href{https://garfieldpp.web.cern.ch/garfieldpp}{https://garfieldpp.web.cern.ch/garfieldpp}}
  (2023) }.

\bibitem{Riegler2002}
W.~Riegler, \emph{{Induced signals in resistive plate chambers}},
  \href{https://doi.org/10.1016/S0168-9002(02)01169-5}{\emph{Nuclear
  Instruments and Methods in Physics Research Section A: Accelerators,
  Spectrometers, Detectors and Associated Equipment} {\bfseries 491} (2002)
  258}.

\bibitem{Riegler2004}
W.~Riegler, \emph{{Extended theorems for signal induction in particle detectors
  VCI 2004}},
  \href{https://doi.org/10.1016/S0168-9002(04)01656-0}{\emph{Nuclear
  Instruments and Methods in Physics Research, Section A: Accelerators,
  Spectrometers, Detectors and Associated Equipment} {\bfseries 535} (2004)
  287}.

\bibitem{Riegler2019}
W.~Riegler, \emph{{An application of extensions of the Ramo–Shockley theorem
  to signals in silicon sensors}},
  \href{https://doi.org/10.1016/J.NIMA.2019.06.056}{\emph{Nuclear Instruments
  and Methods in Physics Research, Section A: Accelerators, Spectrometers,
  Detectors and Associated Equipment} {\bfseries 940} (2019) 453}
  [\href{https://arxiv.org/abs/1812.07570}{{\ttfamily 1812.07570}}].

\bibitem{Janssens2022}
D.~Janssens et~al., \emph{Induced signals in particle detectors with resistive
  elements: Numerically modeling novel structures (vci 2022)},
  \href{https://doi.org/10.1016/J.NIMA.2022.167227}{\emph{Nuclear Instruments
  and Methods in Physics Research Section A: Accelerators, Spectrometers,
  Detectors and Associated Equipment} {\bfseries 1040} (2022) 167227}.

\bibitem{Alexopoulos2011}
T.~Alexopoulos et~al., \emph{A spark-resistant bulk-micromegas chamber for
  high-rate applications},
  \href{https://doi.org/10.1016/J.NIMA.2011.03.025}{\emph{Nuclear Instruments
  and Methods in Physics Research Section A: Accelerators, Spectrometers,
  Detectors and Associated Equipment} {\bfseries 640} (2011) 110}.

\bibitem{Byszewski2012}
M.~Byszewski and J.~Wotschack, \emph{Resistive-strips micromegas detectors with
  two-dimensional readout},
  \href{https://doi.org/10.1088/1748-0221/7/02/C02060}{\emph{Journal of
  Instrumentation} {\bfseries 7} (2012) C02060}.

\bibitem{Riegler2016}
W.~Riegler, \emph{{Electric fields, weighting fields, signals and charge
  diffusion in detectors including resistive materials}},
  \href{https://doi.org/10.1088/1748-0221/11/11/P11002}{\emph{Journal of
  Instrumentation} {\bfseries 11} (2016) }
  [\href{https://arxiv.org/abs/1602.07949v1}{{\ttfamily 1602.07949v1}}].

\bibitem{COMSOL}
COMSOL\textsuperscript{\textregistered}, \emph{Software for multiphysics
  simulation}, {\emph{\href{https://www.comsol.ch}{https://www.comsol.ch}}
  (2023) }.

\bibitem{Smirnov2005}
I.B.~Smirnov, \emph{Modeling of ionization produced by fast charged particles
  in gases}, \href{https://doi.org/10.1016/J.NIMA.2005.08.064}{\emph{Nuclear
  Instruments and Methods in Physics Research Section A: Accelerators,
  Spectrometers, Detectors and Associated Equipment} {\bfseries 554} (2005)
  474}.

\bibitem{MAGBOLTZ}
S.~Biagi, \emph{{MAGBOLTZ+}},
  {\emph{\href{https://magboltz.web.cern.ch/magboltz/}{https://magboltz.web.cern.ch/magboltz/}}
  (2023) }.

\bibitem{Bortfeldt2018}
J.~Bortfeldt et~al., \emph{Picosec: Charged particle timing at sub-25
  picosecond precision with a micromegas based detector},
  \href{https://doi.org/10.1016/J.NIMA.2018.04.033}{\emph{Nuclear Instruments
  and Methods in Physics Research Section A: Accelerators, Spectrometers,
  Detectors and Associated Equipment} {\bfseries 903} (2018) 317}.

\bibitem{Rolandi2008}
L.~Rolandi, W.~Riegler and W.~Blum, \emph{Particle Detection with Drift
  Chambers}, Springer Berlin Heidelberg, 2~ed. (2008),
  \href{https://doi.org/10.1007/978-3-540-76684-1}{10.1007/978-3-540-76684-1}.

\bibitem{Jones1999}
L.L.~Jones et~al., \emph{The apv25 deep submicron readont chip for cms
  detectors}, \href{https://doi.org/10.5170/CERN-1999-009.162}{\emph{Conf.
  Proc. C 9909201} (1999) 162}.

\bibitem{Raymond2000}
M.~Raymond et~al., \emph{The apv25 0.25 $\mu$m cmos readout chip for the cms
  tracker}, \href{https://doi.org/10.1109/NSSMIC.2000.949881}{\emph{IEEE
  Nuclear Science Symposium and Medical Imaging Conference} {\bfseries 2}
  (2000) }.

\end{thebibliography}\endgroup

\end{document}